\documentclass[letterpaper]{article}

\usepackage{natbib,alifeconf}  

\usepackage[utf8]{inputenc} 
\usepackage[T1]{fontenc}    
\usepackage{hyperref}       
\usepackage{url}            
\usepackage{booktabs}       
\usepackage{amsfonts}       
\usepackage{nicefrac}       
\usepackage{microtype}      
\usepackage{lipsum}
\usepackage{fancyhdr}       
\usepackage{graphicx}       
\usepackage{amsmath}
\usepackage{amssymb}
\usepackage{mathtools}
\usepackage{amsthm}

\usepackage{mathtools}
\usepackage{textcomp}

\newtheorem{prop}{Proposition}
\newtheorem{remark}{Remark}

\newtheorem{definition}{Definition}

\title{Capital as Artificial Intelligence}

\author{
    Cesare Carissimo$^{1,*}$, \and
    Marcin Korecki$^{1,*}$, 
    \mbox{}\\
    $^1$ETH Zurich, Switzerland \\
    $^*$Equal Contribution \\
    cesare.carissimo@gess.ethz.ch
} 

\begin{document}
\maketitle

\begin{abstract}
    We gather many perspectives on Capital and synthesize their commonalities. We provide a characterization of Capital as a historical agential system and propose a model of Capital using tools from computer science. Our model consists of propositions which, if satisfied by a specific grounding, constitute a valid model of Capital. We clarify the manners in which Capital can evolve. We claim that, when its evolution is driven by quantitative optimization processes, Capital can possess qualities of Artificial Intelligence. We find that Capital may not uniquely represent meaning, in the same way that optimization is not intentionally meaningful. We find that Artificial Intelligences like modern day Large Language Models are a part of Capital. We link our readers to a web-interface where they can interact with a part of Capital.
\end{abstract}






In this paper we wish to study a component of the economic system which most of us are a part of and which most of us reinforce. In so far as it represents a system much larger than any individual, we wish to study it for its emergent properties in a holistic manner. It possesses qualities of the artificial as it lives in quantities, and its creation is inextricable from the work of man. It also possesses qualities of the emergent which produce behaviour that can appear agential and intelligent. Alongside current reductionist perspectives, we believe we must also study this system as if it were an organism, a sort of artificial life.

The digital revolution of the 20th and 21st centuries has brought a great wealth of novel computational methodologies. Many of these methods yield themselves to applications beyond their initial computational domain. They can be readily used to study systems which are quantifiable. At the same time, a risk of unlimited application of computationalism to subjects that are not readily quantifiable emerges. Indeed, it might be argued that some systems may not be quantifiable without a significant loss, and perhaps more dangerously whose quantification might yield an illusion of understanding. Economics, however, is a field that has always dealt with quantities. More so its main subject might even be said to arrive already quantified\footnote{\emph{`Every non-barter commodity exchange, or non-financial transaction, catalyzes a flow, by pricing it. The ‘real economy’ is thus automatically captured, by a monetary code, even when – as is overwhelmingly typical – this quantification through signs is never aggregated. Since
money systems install an intrinsic economic intelligence (preceding all reflective theorization), economics – uniquely among the sciences – inherits a field of objectivity whose arithmetization has already taken place.'} \citep{land2018crypto}. While economics might not limit itself to the realm of quantities (by also studying the social aspects) its focus is on studying a system whose inherent feature is that of quantification.}. There are many names that have been given to this subject but we will refer to it as Capital, and it is a large, spreading, and complex `beast'.

While the word Capital brings an intuitive association with the economic processes its precise definition is elusive at best. Capital appears to have many facets and according to some theories can itself be broken down into subcategories. The challenges of defining the term, which appears at the core of any economic endeavor, have by no means been lost on its researchers \citep{hayek2019pure}. In classical economics Capital is divided into fixed and circulating capital. The former constituting the goods that are not consumed in the production process and the latter being those goods that are consumed in production \citep{adam2016wealth}. According to the neoclassical synthesis, that combines the Keynesian economics with the neoclassical economics, Capital refers to those durable produced goods that have potential to be used for further production \citep{samuelson2012economia}. The Austrian School, on the other hand, as exemplified by the work of F.A. Hayek, would assign the name Capital to `the total stock of the non-permanent factors of production' \citep{hayek2019pure}. According to Marx, Capital can be viewed as always seeking to create value and surplus-value. As this surplus-value is created via labour, Capital takes a more abstract form than the previously mentioned definitions\footnote{\emph{`Capital is dead labour, that, vampire-like, only lives by sucking living labour, and lives the more, the more labour it sucks.'} and \emph{`The circulation of money as capital is, on the contrary, an end in itself, for the expansion of value takes place only within this constantly renewed movement. The circulation of capital has therefore no limits' \citep{marx2004capital}} present the Marxist sentiment towards Capital. Note the highly anthropomorphic language and an implicit, albeit perhaps metaphorical, endowing of Capital with agency.}. Finally, more provocative definitions of Capital exist and are exemplified by, among others, Nick Land, who sees Capital as limitless dehumanizing consumption of life and intelligence that creates new forms of these and is essentially an auto-promoting feedback loop. Land's perspective would often seemingly equate Capital with life\footnote{\emph{'As capital ’evolves’, the increasingly absurd rationalization of production-for-profit peels away like a cheap veneer from the positive-feedback detonation of production-for-production'} \citep{land1988kant} and \emph{`Life continues, and capitalism does life in a way it has never been done before'} \citep{land2011critique} are some of Land's positions on the topic, which take the organic analogies of Marx even further.}. Further, he characterises Capital as an end in itself rather than as means to an end.

A seemingly common property of most pertinent definitions is that Capital represents value. In particular Capital is that which has a potential to produce more Capital\footnote{The autopoietic undertones should not be missed here \citep{maturana1991autopoiesis, luhmann2012theory}.}. Thus, Capital could be summed up as that value which might produce more value. In this work we refer to Capital with uppercase `C' to differentiate it from the many definitions of capital such as intellectual capital or natural capital known in modern economics. On the other hand, in referring to the features of our model we will use a lowercase `c'. At the end of this section we will attempt a synthesizing of the above definitions taken from traditions between which considerable tensions have accumulated.

As an abstract value that breeds abstract value, Capital and its medium of quantification have grown within, or maybe over, the social landscape. Whether it serves humanity or subjugates it has not been agreed on and both perspectives have been argued for many times. Nevertheless, in manufacturing pure value Capital is driven by a vast number of humans, their technique and labour. As such it has been interpreted as an expression of a spontaneous order that emerges from all the aims of individuals that take part in it \citep{hayek2011law}. Instead of adopting this perspective uncritically we will modify it while accepting that Capital is indeed best understood as a complex system. 

While Capital, as arising from the social, is at its core the emergent of a complex system, it is not directly clear that it embodies the perfect aggregation of preferences of all individuals that take part in it. As it is, any complex system is essentially more than the sum of the constituting parts \citep{heylighen2006complexity}. This manifests in practice in the growing frustration of realising that as individuals we have limited control over Capital. Naturally, the fact that a single individual is not able to influence the entire system does not imply that the system does not take into account that individual's aims. However, within Capital we seem to observe situations, where a vast majority of individuals seem to agree on certain abstract goals and yet these goals do not appear to be readily reflected in the emergent goals of Capital. The environmental crisis or the persistent continuation of the military industrial complex exemplify this property. The majority of people seem to agree that the environment should be protected and that wars should not be fought and yet Capital continues to grow by ravaging nature and dealing in death. And perhaps that is not at all surprising as the uncontrollable character of Capital arises from its identity as an emergent. Indeed Capital can also grow through investments that improve nature, and trade — and thus capital growth — flourishes more in peacetime than wartime. Thus, something akin to a will of its own arises in Capital that can be neither ruled nor explained in terms of its constituents\footnote{It is this independence, among other things, that breeds the curious form of resentment, disengagement and alienation characteristic of our modern days. Within Capital it is not only the individual who is disempowered but also aggregations of individuals.}.

The complexity perspective, arising from cybernetics, offers a valuable view on Capital, which through its lens can be understood as a positive reinforcement feedback loop \citep{mirowski2002machine}. As such Capital is that, which in its continuation constantly yields more of itself. This agrees with the traditional definitions that would see Capital as the value that has a potential to produce more of itself. Indeed, many economic processes have already been identified as being highly self-reinforcing \cite{arthur2018self}. In understanding Capital as a system we can extend the function of its feedback character to include the curious property, analogous to autopoeisis, that allows it to produce and maintain its own structure\footnote{\emph{"Capitalism" [...], designates the sovereign self-escalation of an innovative entity, defined only by the practical relation of auto-promotion it establishes with – and through – itself. What it is, in itself, is more than itself. Growth is its essence.'} \citep{land2018crypto}}. This self-productive maintenance, however, might also, at times, fail leading to a degeneration of the system in the form of an economic crisis. The Capital then could be seen as undergoing phases akin to growing and declining or even living and dying. An aversion towards letting the Capital fail has remained a popular sentiment in many circles \citep{mirowski2014never} but it could be argued that letting the system `die' could allow it to learn and improve \citep{korecki2023artificial}. Moreover, these common analogies of living and dying further support the potential of an organic perspective on Capital.  

Historically, the emergence of surplus value might be associated with the beginning of civilisation (10000BC), where a surplus of goods would be created by applying a novel agrarian technique\footnote{We use technique in the sense proposed by Jacques Ellul. \emph{`Technique is the totality of methods, rationally arrived at and having absolute efficiency (for a given stage of development) in every field of human activity'} \citep{ellul2021technological}.}. The pre-agrarian hunter-gatherer societies did not have a concept of wage labour because they did not have currency. Labour might have been rewarded with concrete services or goods and commodities may have been exchanged directly. Nevertheless, the lack of a universal medium of commensurabilisation made it impossible to either accumulate or consistently quantify value. But through currency the surplus goods could be converted and stored. Assuming the price of producing the goods was lower than the selling price, surplus value would be created. As such the historical emergence of Capital appears intrinsically linked both to the surplus goods being created and the advent of currency as an accepted medium of exchange.

It appears, though, that Capital has changed significantly since those bygone days of its infancy. As time passed, it has become more and more abstracted away from the direct realities of human experience (financial instruments such as derivatives or complex insurance plans serve as examples of the growing abstraction). It has also expanded greatly to include ever more things in its network of commensurability. The leaps in `territory' that Capital covers appear correlated with the paradigm shifts in technique\footnote{In the later parts of this work we argue for an analogy between Capital and autonomous technique which might perhaps add to explaining this apparent historical correlation.}, such as the already mentioned agrarian revolution or more recently the industrial revolution \citep{braudel1992civilization}.

The definition of Capital that we will use in what follows does not aim to add yet another formulation to the already crowded landscape. We posit a minimal and functional definition that encapsulates the characteristics of Capital as identified above, taking into account its historical context and the way it is experienced by individuals. 

\begin{definition}
    Capital is a historical agential system that arises and continues embedded in the complexity of a social sphere. As a system it is embodied in quantified value and that which pursues its maximization.
\end{definition}

\section*{Capital as a Historical Agential System}

To construct our model of capital we inherit useful definitions from Continual Reinforcement Learning \citep{abeldefinition}. Capital then extends the definitions to suit the needs of Capital: a system which at its core extends as it is extended.

We start with an \textit{agent-environment interface} \citep{abeldefinition} defined by the tuple $(\mathcal{A}, \mathcal{O})$ of actions and observations. The set of all possible histories given $(\mathcal{A}, \mathcal{O})$ is defined as:
\begin{equation*}
    \mathcal{H} = \bigcup_{t=0}^{\infty} (\mathcal{A} \times \mathcal{O})^{t}
\end{equation*}
A single history $h$ is a sequence of observation action tuples $(o,a)$, and is an element of $\mathcal{H}$. The environment is defined as a function which maps histories and actions to a probability distribution over observations: $e:\mathcal{H}\times\mathcal{A} \rightarrow \Delta(\mathcal{O})$. We define a reward function which maps an observation and an action to a real number: $r:\mathcal{O}\times\mathcal{A}\rightarrow \mathbb{R}$. We denote agents by $\lambda$, and agents map histories to a probability distribution over actions: $\lambda: \mathcal{H} \rightarrow \Delta(\mathcal{A})$.

The \textit{realizable histories} \citep{abeldefinition} of a given agent-environment pair $(\lambda, e)$, define the set of histories of any length that can occur with non-zero probability from the interaction of $\lambda$ and $e$,
\begin{equation*}
    \mathcal{H}^{\lambda, e} = \bigcup_{t=0}^{\infty} \bigg\{ h_t \in \mathbf{H}_t \: : \: 
    \prod_{k=0}^{t-1} e(o_{k+1} | h_k, a_k) \lambda(a_k | h_k) > 0 \bigg\}.
\end{equation*}

The \textit{realizable history suffixes} \citep{abeldefinition} of a given $(\lambda, e)$, relative to a history prefix $h \in \mathcal{H}^{\lambda, e}$, define the set of histories that, when concatenated with prefix $h$, remain realizable,
\begin{equation*}
    \mathcal{H}^{\lambda, e}_h = \bar{\mathcal{H}}_h = \{ h' \in \mathcal{H} \: : \: hh' \in \mathcal{H}^{\lambda, e}\}.
\end{equation*}
\begin{prop}
\label{prop:historical}
    Capital is a historical agential system.
\end{prop}
We take the perspective of Capital as a historical process in that it is instantiated in time and shaped by its sequential history. We characterize its history as the past observations and the actions that were taken at those observations. In this manner, a minimal instance of Capital is: an agent $\lambda$ and an environment $e$ which together establish the set of realizable histories $\mathcal{H}^{\lambda, e}_h$, which are analogously the future possibilities of its becoming. The actions $a$ and observations $o$ are afforded to the agent by Capital so that the interactions between the agent and the environment are always mediated by Capital.

The environment represents the entire space in which Capital is embedded (the Earth, the Universe). Across that environment Capital extends its own medium through which the observations are accessed. Capital perceives the environment through its observations, and the two need not be in a one-to-one relationship.
\newline


\begin{prop}
\label{prop:discrete}
    Capital is fundamentally discretized (digital) into units of capital.
\end{prop}

The medium of Capital is quantified value. In the history of societies this quantification has come in the form of money and currency. By quantifying the value of things we embed them in a space where they can be compared\footnote{These things, whose value we quantify, might have subjective or qualitative value to us but as part of Capital they must exist as quantities.}. Thus the universal medium of commensurability is established, in which Capital operates.

Historically all currencies have quantified value in a countable way. Intervals on uncountable infinities all contain the same infinity, and thus are not meaningfully commensurable, they all are the same size infinity. 
Since commensurable quantification can only be done on countable sets the units of quantified value (such as currency) are discrete. The digitalness follows from the ability to represent all countable sets as 0's and 1's.

Therefore, we define a smallest unit of capital, $0 < \mbox{\textcent} << 1$, much like there is a smallest unit of Euro, Dollars and Bitcoin. Every capital reward $r$ is expressable as a multiple of \textcent: $r = k \mbox{\textcent}, k \in \mathrm{N}$. Accumulated capital in time can be expressed as the sum over all the capital rewards:
\begin{equation*}
    G_t = \sum_{i = 0}^t r_i
\end{equation*}
\begin{prop}
\label{prop:units}
    Each unit of capital affords observations and actions.
\end{prop}

In its fully discretized form, we can understand the units of capital as deployed in the environment where they afford observations of the environment and possible actions to the agents. At every timestep $t$, an observation for the system is constituted of an observation for each unit of capital,
\begin{equation*}
    \bar{o} = \{o_1, o_2, \dots, o_m\}
\end{equation*}
and similarly each unit of capital affords actions from the observed state,
\begin{equation*}
    \bar{a} = \{a_1, a_2, \dots, a_m\}
\end{equation*}
where the number of observations and actions in aggregate, $m$, depends on the accumulated capital $m = kG$. 


Units of capital can afford observations and actions to the agents of capital (e.g. by modifying their observations or available actions), which could take the form of individuals and groups, like citizens, companies, and nations. The actions available to the units from their respective observations \emph{may generate more units}. We develop on the agents of capital in the following section.
\newline

\begin{prop}
\label{prop:actions_time}
    Actions are time dependent.
\end{prop}

It follows from \autoref{prop:actions_observations} and \autoref{prop:observations_time} that there may exist actions from observations at a point in time which did not exist at another point in time.
\begin{align*}
    & \exists (t, \tau, a) \text{ where } t < \tau \: : \\ 
    & \: \forall (o, a') \in \mathcal{H}_t, \: a' \neq a \:\: \cap \:\: \exists (o, a') \in \mathcal{H}_\tau, \: a' = a
\end{align*}

\begin{prop}
\label{prop:actions_observations}
    Actions are observation dependent.
\end{prop}
Not all actions are possible from different observations. There is an inherent local sensitivity to the units of capital. Where the units of capital are in the environment determines the set of actions available to them.
\newline
\begin{equation*}
    \exists o, o' \: : \: o \xrightarrow[]{a} \: \cap \: o' \not{\xrightarrow[]{a}}
\end{equation*}

\begin{prop}
\label{prop:generated}
    Actions can generate new units of capital. A generated unit of capital observes the next observation.
\end{prop}
As a result of taking an action new units of capital may be spawned. Not all actions lead to new units of capital being generated. This corresponds to the inherent property of Capital that is seeking to produce more of itself (surplus value). 
Given a unit of capital with observation $o$, taking action $a$, generating units of capital $r$ and observing a \textit{next} observation $o'$, resulting in the transition tuple $(o, a, r, o')$, the generated units of capital observe the new observation $o'$.
\newline

\begin{prop}
    Capital can be partitioned.
\end{prop}
Capital may act both independently and collectively. As formulated, each unit of capital affords observations and actions. Capital can be partitioned into sets of capital units where capital units in a partition $U$ may be conditionally dependent. A partition $U$ contains multiple capital units, $U = {u, \dots, x}$ where $u,x \in \mathrm{N}$. The partitioning of capital units influences the agency that agents of capital have in Capital. Intuitively, a fine partitioning can reduce the extent of agency, while a coarser partitioning can increase the extent of agency. The agency of a partition is the total amount of units of capital in the partition. We refer to the complete partitioning of Capital as $P$, which is set of all the partitions of capital.

A specific partitioning of Capital $U$ affords the joint probability distribution of the units of capital in $U$. The total capital of a partition is $G(U) = \sum_{u \in U} u$. The partition affords a joint observation space $\bar{o}_U = \{o\}_{o \in U} \in \bar{o}$. The history of a partition $U$ is then defined as an enlarged history $h_{U}$, where each element $h_t$ in the complete history $h_U$ is a tuple of the observations and actions afforded by all the units of capital in $U$: ${(o_{t, x}, a_{t, x})}_{x \in U}$.
\newline

\begin{prop}
\label{prop:nonergodic}
    The observation space is non-ergodic.
\end{prop}

There might be states in the observation space that become unreachable due to the internal dynamics of the environment or the actions of the units of capital. This implies the potential non-reversibility of some actions.  
\begin{align*}
    & \exists o \in \mathcal{O}, \exists a \in \mathcal{A} \:\text{where}\: o \xrightarrow{a} o', h \xrightarrow[]{a, o'} h' \: : \\ 
    & \: \forall a' \in \mathcal{A}, (a', o) \notin \mathcal{H}^{\lambda, e}_{h'}
\end{align*}

\begin{prop}
\label{prop:observations_time}
    The observation space changes in time.
\end{prop}

There may exist different observations at different points in time, which did not exist at other points in time. This can be due to the internal dynamics of the environment and the actions of the units of capital. For example, an exogenous natural catastrophe beyond the control of Capital will alter the observable environment independently of the actions of the units of capital. Conversely, the actions of the units of capital can create new things which themselves become parts of the observations (e.g. a computer allowing higher dimensional observations of the environment by through computation).
\begin{align*}
    &\exists (t, \tau, o) \text{ where } t < \tau \: : \\
    & \: \forall (o', a) \in \mathcal{H}_t, \: o' \neq o \:\: \cap \:\: \exists (o', a) \in \mathcal{H}_\tau, \: o' = o
\end{align*}

\autoref{prop:nonergodic} and \autoref{prop:observations_time} essentially represent the properties of an open dynamic system, such as the universe, in which Capital is embedded.
\newline

\section*{The Agents of Capital}
\label{sec:agents}

\begin{figure}
    \centering
    \includegraphics[width=1\linewidth]{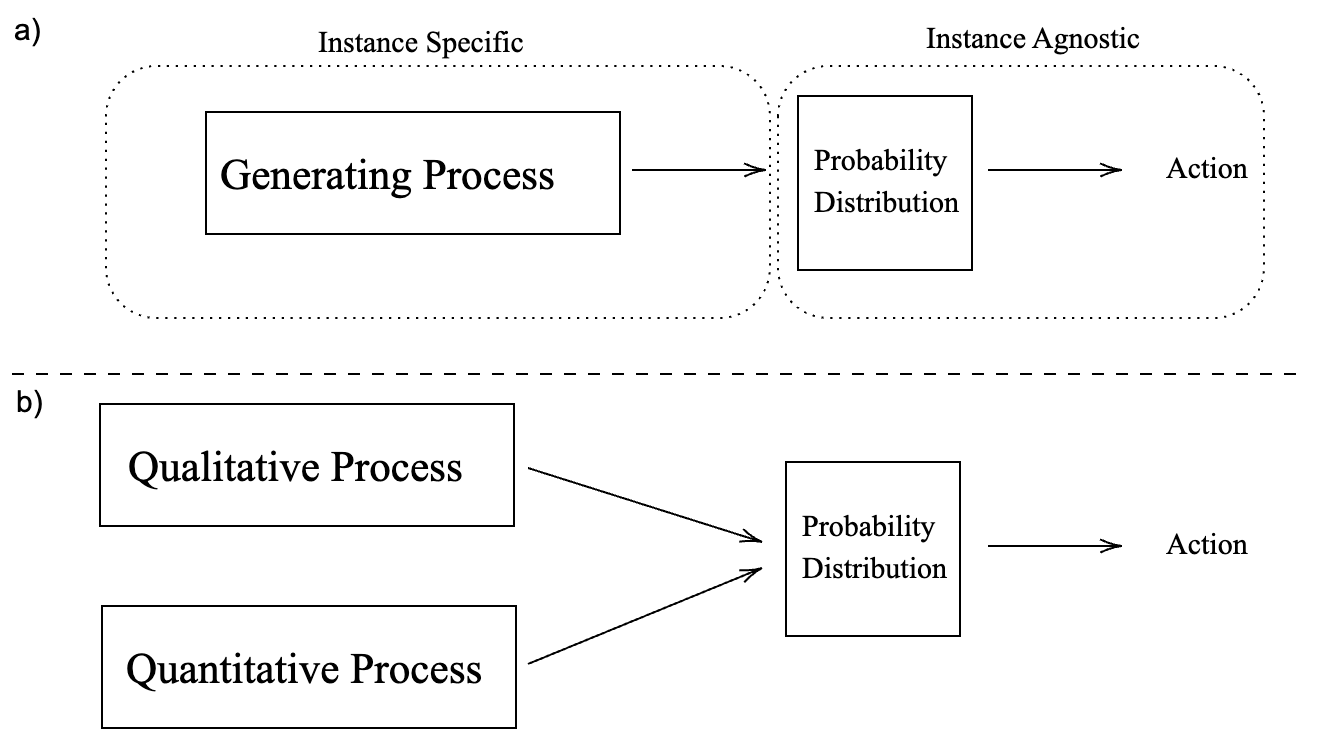}
    \caption{a) The relationship between generating processes and probability distributions. The generating process determines the probability distribution which is sampled from to determine the action. b) Generating processes can be either qualitative or quantitative.}
    \label{fig:generating-process}
\end{figure}

As we have proposed, Capital is composed of units of capital and agents of capital that determine the probability distributions over actions for units of capital given their histories. More generally, a capital agent can determine the probability distribution for a partition of capital $U$. Such a capital agent can correlate the actions of the capital units such that their probability distributions are no longer independent.

\begin{prop}
    The goal of a capital agent is to increase itself by pursuing capital accumulation as expressed by the amount of units of capital in its partition.
\end{prop}

One way to formally define this maximization is through the cumulative sum of discounted future rewards, for a given discount factor $0 \leq \gamma \leq 1$ (following in the reinforcement learning tradition), starting from the current time $\tau$:
\begin{equation*}
    \pi_{U} = \sum_{t=\tau}^{\infty} \sum_{x\in U} \gamma^t r_x^t,
\end{equation*}
where $r_x^t$ is the capital reward obtained by capital unit $x$ at time $t$. At every time $t$ the goal of a capital agent is to maximize future capital accumulation. The discount factor $\gamma$ establishes the extent to which the capital agent over partition $U$ denoted $\lambda_U$ is myopic or far-sighted; the extent to which the agent values short term or long term rewards.

\begin{remark}
    Maximizing capital is ill-defined for an open system.
\end{remark}

The goal of capital maximization can be pursued with optimization techniques, but there is no guarantee that optimization will reach the maximum in the system as presented. This is due to the fact that the observation and action spaces can change in time and that the observation space is non-ergodic. Nonetheless, a capital agent can still pursue capital accumulation.


\begin{remark}
    The evolution of Capital is modelled as a probability distribution $\mathbf{Y}$ which is the joint distribution of all the units of capital. Given a partitioning of capital $P$, each agent $\lambda_U, U \in P$ establishes the joint distribution $\mathbf{Y}_U$ over the capital units $x \in U$.
\end{remark}

Our model propositions define a template which is agnostic to the generating processes (see \autoref{fig:generating-process}) of the probability distributions of capital units, and can then be used to model Capital. A grounding of our model of Capital adheres to our propositions and additionally inherits the generating processes of the probability distributions, as in \autoref{fig:generating-process}(a). Importantly, the generating processes determine the probability distributions for the agents which control partitions of capital.

As humans we inhabit a specific grounding of Capital and our direct experiences reveal broad categorizations of generating processes. Humans are arguably responsible for modern day Capital, fuelled by the viral desire to quantify and accumulate it; in so doing humans are agents of capital which serve and reinforce Capital. They are not, however, the sole agents. Communities, companies, nations and parallel economic systems are themselves agents that can control Capital, and they too are agents of capital in so far as they pursue its accumulation. Specific instances of these groups can be modelled as partitions of capital. For agents of capital we identify two general kinds of generating processes: qualitative processes and quantitative processes (\autoref{fig:generating-process}(b)). In this split, we implicitly assume there exist qualities which cannot be quantified, and we also assume that such qualities are within the realm of experience of human individuals as they possess an own contingency \footnote{Such a distinction can be linked to the \textit{Lebenswelt} or "Lifeworld" concept popular of the German philosophers of the 20th Century \citep{husserl1970crisis}: the realm of experience, the state of affairs as experienced, that which precedes epistemological analysis, and in our understanding that which precedes quantification.}. We posit these methods lie beyond quantification and are not reducible to it\footnote{An example of a qualitative generative process could be an ethical value system held by a given human and used to influence their actions within the system of Capital. We stand by the argumentation by which quantitative analysis cannot be used to take normative decisions \citep{korecki2023artificial}.}.

On the other hand, quantitative processes are well suited to mathematical tools. In fact, any pursuit of capital accumulation with quantitative processes lies within the realm of optimization. We put forward the following remark:

\begin{remark}
    Any use of Quantitative Processes to generate probability distributions are subsumed by the methods of optimization.
\end{remark}

Our remark stands to indicate that within the realm of processes that operate on quantities, any process can be modelled using mathematical tools such as optimization. Later we extend this discussion to understand the manners in which Capital can be artificially intelligent.

\section*{The Entropy of Capital}

Using our model we can reason about the entropy of Capital. The information entropy of a discrete random variable $X$, with support $\mathcal{X}$, and distributed according to $p:\mathcal{X}\rightarrow [0,1]$ is,

\begin{equation*}
    H(X) := - \sum_{x\in \mathcal{X}} p(x) \log p(x).
\end{equation*}

We take this definition of information entropy and extend it to describe the entropy of Capital. First, we must clarify that we will define its information entropy for given timesteps $t$. Then, we identify the random variables of capital as the discrete distributions over actions given observations induced by the agents of capital. For each unit of capital $\mbox{\textcent}_i$ we can compute its corresponding entropy, given a history of that unit $h_i$, as the information entropy of a random variable $Y_i$ with support $\mathcal{O}(i)$, and distributed according to $q:\mathcal{O}(i) \rightarrow [0,1]$,

\begin{equation*}
    q(o_i) = \sum_{a\in \mathcal{A}(i)} \lambda(a|h_i) \times e(o_i|h_i,a),
\end{equation*}

where $\mathcal{A}(i)$ is the set of actions available to $i$. Thus expressed, $q$ provides the distribution of next observations with which we can calculate the entropy of each unit of capital:

\begin{equation*}
    H(Y_i) := - \sum_{o\in \mathcal{Y}_i} q(o) \log q(o).
\end{equation*}

Next, we can express the entropy of Capital as the entropy of the joint probability distribution:

\begin{align*}
    H(\mathbf{Y}) & = H(Y_1, \dots, Y_m) \\
    & = - \sum_{o_1\in \mathcal{Y}_1} \dots \sum_{o_m\in \mathcal{Y}_m} q(o_1, \dots, o_m) \log( q(o_1, \dots, o_m) ).
\end{align*}

Establishing conditional dependence between capital units is the main way in which agents of capital can reduce the entropy of Capital.
\newline

The language used by Marx and many others is populated by metaphors that present Capital as an organism. Nick Land takes it further by seemingly equating Capital with life. As such Capital is a system that could be of great interest to the Artificial Life (alife) perspective. Life is often characterized also by its ability to minimize entropy in a universe whose entropy is always increasing. Entropy has also been used to analyze capitalism \citep{biel2011entropy}, though in a reservedly qualitative and discursive manner. By extending a definition of entropy, whose minimization has been used as an indicator of what we define as life, to Capital we make it possible to study the metaphors and links established by the aforementioned authors in a more formal way. 




\section*{The Artificial Intelligence in Capital}

In this section we wish to tease out the profound relations that exist between Capital and Artificial Intelligence (AI). The relation goes both ways. The first direction is understanding AI as an agent of capital. The second direction is understanding Capital as AI.

To be clear, by AI we mean deep machine learning models trained via mathematical optimization. This categorization fits any of the impactful AIs that we interact with today, from Large Language Models like Chat GPT to content recommender systems used by Instagram, Youtube and Google. All of these systems are massive parameter architectures that are fitted through mathematical optimization tools.

\paragraph{Artificial Intelligence is an agent of Capital}


Through our model we have embedded Capital in a formal language which comes from the field of reinforcement learning and mathematical optimization found in control theory and economics. AI as we have it in the year $2024$ is also built in this language. Thus, our model of Capital takes on a form `digestible' by AI. In concrete terms AI can readily be deployed to interact with Capital as an agent, given a well defined agent-environment interface and adherence to our propositions. 

\begin{figure}
    \centering
    \includegraphics[width=\linewidth]{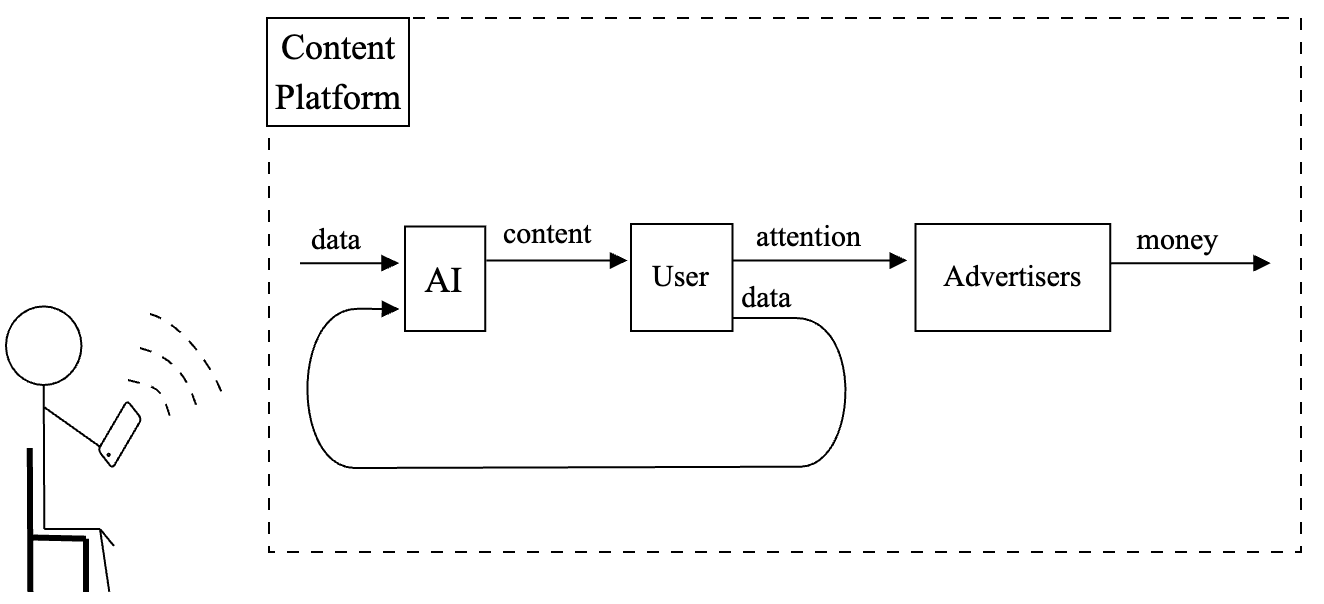}
    \caption{A content platform takes users' data to produce recommendations, and in the background uses an AI to process the quantified value, and sell the attention to an advertiser. The recommender algorithm may be interpreted as an agent of capital with the data constituting the observation and the recommednations being the actions.}
    \label{fig:recommender}
\end{figure}

Consider an example of a recommender algorithm, which is nothing else than a massive deep learning model. It is fed the data on the users' behavior and preferences (quantified) and outputs content recommendations directly to the users (\autoref{fig:recommender}). The recommendations are optimized such that they maximise users' engagement and time spent on a given platform. The data fed into the model is clearly valuable (because it can be used to produce value), the output is also clearly valuable (because it drives engagement, which drives revenue). Here we note that the AI is fed quantified value and outputs more quantified value, thus, acting, according to our terminology, as an agent of capital. Additionally, AI processes quantities that are much greater than what we could process ourselves.

AI as an active agent of capital may take the form of recommender algorithms but also algorithmic trading or LLMs and other similarly automated solutions. This is possible as the generative processes of the probability distributions of Capital can be purely quantitative in how they specify the pursuit of accumulation. Thus, AI, being essentially an optimization process \citep{carissimo2023limits} fits perfectly within our model of Capital\footnote{Consider a radical extension of the idea that AI agents can be agents of Capital, where the Capital is no longer dependent on humanity for its continued functioning. A fully automated AI driven Capital system is not an impossibility. The difference between humans and AI as agents of Capital is that humans are capable of employing both the qualitative and quantitative generative processes, while AI agents are limited only to the quantitative.}.


We have argued that the inputs of the AI are quantified value, and in that sense can be understood as units of Capital. Then, by their very nature, AI can be designed to produce the `best possible' processing of the quantified value through quantitative processes. In so doing they pursue Capital accumulation. AI's are therefore the best possible agents of capital that humans have been able to create. In creating AI, humans have also been very good agents of capital. 

\begin{remark}
    We can see AI as a specific subset of Capital, which fits into the quantitative generative processes of the probability distributions of the units of capital, and as such is an agent of capital. We can see AI as one of the ways in which Capital has developed to process large quantities of information to create more value.
\end{remark}


\paragraph{Capital as Artificial Intelligence}

Next we will argue that we can learn about Capital by understanding AI, and that there are ways in which Capital is an AI. As the first parallel we indicate that, much like AI can process far more quantified value than humans are capable of, the economy and markets in the economy are also able to process much more quantified value than any individual economic agent. 


Capital can be understood as an optimization, in the very least as an ongoing attempt of optimization, when the generating processes are quantitative. We have also established that Capital processes quantified value in amounts much larger than individual human beings can. Together, these two points afford a striking resemblance to AI. The main difference remains that the generating processes of the distributions of capital may be qualitative, which remains outside of the scope of AI. In light of this difference, we make the claim that Capital can be \textit{artificially intelligent} in proportion to the generating processes of the probability distributions of capital which are quantitative.

\begin{remark}
    In so far as the generating processes of the probability distributions of capital are optimization, Capital can be understood as artificially intelligent in its ability to: process much larger quantities of value than its constituents, and pursue goal directed capital accumulation which produces new information in the form of quantified value.
\end{remark}

\section*{A Pragmatic Perspective}

Thus, we have claimed that Capital can be artificially intelligent, and in so far as it is artificially intelligent this has implications for the manners in which we should interpret the value produced by Capital. To do so we must explore "meaning", where by meaning we denote the interpretation of words or information as done by the receiver of the information.

When a Large Language Model (LLM), exemplifying ML and trained in the process of optimization, produces a response to an input prompt we might interpret it as meaningful. We may know that ultimately the LLM is not trying to make sense, it is just outputting a probability distribution it has encoded in its weights, and yet we interpret the things that it says. We can find meaning in the text which an LLM produces knowing very well that there is no intention behind it. When optimizations produce outputs that we could not have produced ourselves, and that we find meaningful, we tend to attribute qualities of intelligence to them, colloquially referring to them as AI. However, AIs only have a contingency that can produce meaning by reflection of the contingencies of the human agents which produce its training data \cite{esposito2017artificial}, because optimizations do not possess a normative dimension required for intent \citep{korecki2024man}. If the output of an AI has meaning it is a reflection of its meaningful inputs.

Then, by composition with our argument that Capital can be artificially intelligent, there is an extent to which Capital driven by quantified optimization processes is as devoid of intent as the output of a Large Language Model. As for AI, Capital driven by quantitative optimization processes can at best reflect the meaning of the inputs as generated by beings with their own contingencies like humans.

\begin{remark}
    In so far as the generating processes of the probability distributions of capital are optimization, we cannot attribute intent to the meaning that we interpret in the quantified value that Capital produces.
\end{remark}

\begin{figure}
    \centering
    \includegraphics[width=1\linewidth]{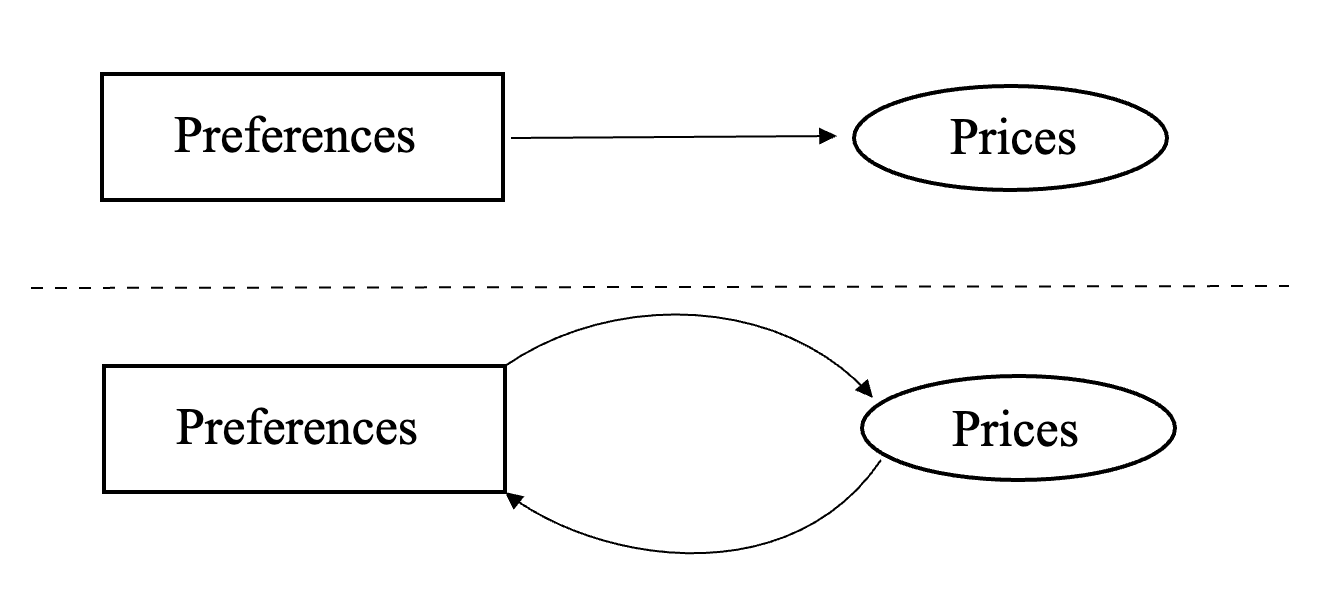}
    \caption{Top: prices are determined by preferences, so the meaning of prices is aggregate preferences. Bottom: prices are determined by preferences, which are influenced by prices. As prices influence preferences, the aggregate preferences revealed by prices are tied to the prices themselves, so the meaning of prices is void in the absence of prices.}
    \label{fig:preferences}
\end{figure}

On the other hand, in interpreting the information of Capital, what is its meaning? Let us look at prices, one of the main quantities produced by Capital. What meaning can we sensibly ascribe to prices? To an economist, prices reflect the preferences of market participants. Even if the relationship between prices is bi-directional (\autoref{fig:preferences}), if a market perfectly reflects preferences, in that the preferences of all capital agents are considered, this produces a unique equilibrium. If this is true then the meaning of prices could be the reflection of the preferences of capital agents, an act of `artificial communication' \cite{esposito2017artificial} achieved through large scale mechanistic apparatus of Capital.

However, consider financial crises, or bubbles, where the sudden change in the value of an object of capital can alter the preferences of people. Such events stretch the fabric of Capital revealing a reality of feedback loops in which it is no possible to disentangle the preferences of capital agents from the price information produced by Capital, and a reality which equilibrium concepts do not describe. The preferences of capital agents as revealed through prices in a market may not be meaningful reflections of static preferences, but rather dynamic preferences which are themselves influenced by the pursuit of Capital accumulation. We cannot discover the `true' preferences of other capital agents through prices, because the preferences are themselves influenced by prices (\autoref{fig:preferences}). We believe this to be a feature which is inherent to the quantification of value, its subsequent optimization, and its enmeshment in a social sphere. 

We are certainly not the first to propose such systemic perspectives on complex social systems. In the late 20th Century Niklas Luhmann can be credited with a most thorough elaboration of a perspective on society as a system which evolves through functional relationships \citep{luhmann2012theory}. This very point was greatly debated by his contemporary Jürgen Habermas who viewed this functionalisation \footnote{This term is central to the perspectives of these philosophers and is used to indicate relationships between objects which can be formalised, inherently linked to quantification and to the broader debate of computationalism still strong in modern society.} of society as an issue which could obfuscate those things that could not be cast in functional terms \citep{moynahan2018habermas}: in our perspective those things which cannot be quantified and optimized. Thus, rather than a full functionalisation of society, our model of Capital carries forward this thematic debate by delineating the functional relationships that exist within the social spheres that are dominated by value, its quantification and its optimization: the sphere of Capital.

Viewing Capital with our model has implications for the ways in which we, as humans, can interact and engage with Capital. We emphasize that if we are interested in meaning and purpose -- qualities characteristic of entities with their own contingencies -- that we cannot rely exclusively on the information produced by Capital, and must instead look beyond Capital. For example, social relations, interactions with other living systems like plants and animals, even interactions with ourselves as composed by trillions of living cells, are not immediately quantified and allow us to take part in a firstly qualitative experience of reality where we can seek meaning. It is important to note how these spheres of social interaction most dear to us are rapidly being quantified and invaded by Capital accumulation in the early 21st Century, for example, through the extractive practices which sell user behavioural data on social networks (dating apps, influencer profiles). This trend has been on the minds of philosophers for centuries.

\section*{Conclusion}

In this paper, we have defined Capital and proposed a model of it. Our model uses tools from computer science to formalize the agential and historic process that Capital embodies. In our model, Capital is described through capital units and agents of capital that pursue capital accumulation. The capital units afford agency for the capital agents that control them. We acknowledge this is a self-referential definition. We claim that it serves the purpose of studying Capital in a holistic manner. Even though we identify units and agents of capital, we do not reduce Capital to either of them: it is modelled as their interaction. Moreover, we extend the definition of information entropy and derive a measure for the entropy of Capital, which can be useful for further studying the system we model and the connections with artificial life. 

After establishing our model we identify the parallel between Capital and AI. AI is an agent of capital which processes quantified value to create more value. Capital can also be understood as AI when the generative processes of the probability distributions of capital follow optimization. Finally, we establish a parallel between the meaning of outputs of AI and the meaning of prices produced by Capital: as the outputs of AI are not intentionally meaningful, the outputs of Capital (e.g. prices) can be interpreted as meaningful without reflecting inherent meaning \footnote{You can find the \href{https://capital-ai.vercel.app/}{link}. to our Capital LLM, which trained on economic texts, stands to serve as a means of interacting with Capital and understanding the parallels between Capital and AI.}

As a further direction, we envision the work on validating the proposed model through the analysis of historical precedents such as economic crises, large-scale wealth transfers and the global and local distributions of Capital. 



\footnotesize
\bibliographystyle{apalike}
\bibliography{references}

\begin{thebibliography}{}

\bibitem[Abel et~al., 2024]{abeldefinition}
Abel, D., Barreto, A., Van~Roy, B., Precup, D., van Hasselt, H.~P., and Singh, S. (2024).
\newblock A definition of continual reinforcement learning.
\newblock {\em Advances in Neural Information Processing Systems}, 36.

\bibitem[Adam, 2016]{adam2016wealth}
Adam, S. (2016).
\newblock {\em The wealth of nations}.
\newblock Aegitas.

\bibitem[Arthur, 2018]{arthur2018self}
Arthur, W.~B. (2018).
\newblock Self-reinforcing mechanisms in economics.
\newblock In {\em The economy as an evolving complex system}, pages 9--31. CRC Press.

\bibitem[Biel, 2011]{biel2011entropy}
Biel, R. (2011).
\newblock The entropy of capitalism.
\newblock In {\em The Entropy of Capitalism}. Brill.

\bibitem[Braudel, 1992]{braudel1992civilization}
Braudel, F. (1992).
\newblock {\em Civilization and capitalism, 15th-18th century, vol. II: The wheels of commerce}, volume~2.
\newblock Univ of California Press.

\bibitem[Carissimo and Korecki, 2023]{carissimo2023limits}
Carissimo, C. and Korecki, M. (2023).
\newblock Limits of optimization.
\newblock {\em Minds and Machines}, pages 1--21.

\bibitem[Ellul, 2021]{ellul2021technological}
Ellul, J. (2021).
\newblock {\em The technological society}.
\newblock Vintage.

\bibitem[Esposito, 2017]{esposito2017artificial}
Esposito, E. (2017).
\newblock Artificial communication? the production of contingency by algorithms.
\newblock {\em Zeitschrift f{\"u}r Soziologie}, 46(4):249--265.

\bibitem[Hayek, 2011]{hayek2011law}
Hayek, F.~A. (2011).
\newblock {\em Law, legislation and liberty, volume 1: Rules and order}, volume~1.
\newblock University of Chicago Press.

\bibitem[Hayek and White, 2019]{hayek2019pure}
Hayek, F.~A. and White, L.~H. (2019).
\newblock {\em The pure theory of capital}.
\newblock Routledge.

\bibitem[Heylighen et~al., 2006]{heylighen2006complexity}
Heylighen, F., Cilliers, P., and Gershenson, C. (2006).
\newblock Complexity and philosophy.
\newblock {\em arXiv preprint cs/0604072}.

\bibitem[Husserl, 1970]{husserl1970crisis}
Husserl, E. (1970).
\newblock {\em The crisis of European sciences and transcendental phenomenology: An introduction to phenomenological philosophy}.
\newblock Northwestern University Press.

\bibitem[Korecki et~al., 2023]{korecki2023artificial}
Korecki, M., Carissimo, C., and Lund, T. (2023).
\newblock artificial death: learning from stories of failure.
\newblock In {\em ALIFE 2023: Ghost in the Machine: Proceedings of the 2023 Artificial Life Conference}. MIT Press.

\bibitem[Korecki et~al., 2024]{korecki2024man}
Korecki, M., K{\"o}stner, G., Martinelli, E., and Carissimo, C. (2024).
\newblock The man behind the curtain: Appropriating fairness in ai.
\newblock {\em Minds and Machines}, 34(1):7.

\bibitem[Land, 1988]{land1988kant}
Land, N. (1988).
\newblock Kant, capital, and the prohibition of incest: A polemical introduction to the configuration of philosophy and modernity.

\bibitem[Land, 2011]{land2011critique}
Land, N. (2011).
\newblock Critique of transcendental miserablism.
\newblock {\em Fanged Noumena: Collected Writings 1987--2007}, pages 623--627.

\bibitem[Land, 2018]{land2018crypto}
Land, N. (2018).
\newblock Crypto-current: An introduction to bitcoin and philosophy.
\newblock {\em Retrieved August}, 22:2020.

\bibitem[Luhmann, 2012]{luhmann2012theory}
Luhmann, N. (2012).
\newblock {\em Theory of society, volume 1}.
\newblock Stanford University Press.

\bibitem[Marx, 2004]{marx2004capital}
Marx, K. (2004).
\newblock {\em Capital: volume I}, volume~1.
\newblock Penguin UK.

\bibitem[Maturana and Varela, 1991]{maturana1991autopoiesis}
Maturana, H.~R. and Varela, F.~J. (1991).
\newblock {\em Autopoiesis and cognition: The realization of the living}, volume~42.
\newblock Springer Science \& Business Media.

\bibitem[Mirowski, 2002]{mirowski2002machine}
Mirowski, P. (2002).
\newblock {\em Machine dreams: Economics becomes a cyborg science}.
\newblock Cambridge University Press.

\bibitem[Mirowski, 2014]{mirowski2014never}
Mirowski, P. (2014).
\newblock {\em Never let a serious crisis go to waste: How neoliberalism survived the financial meltdown}.
\newblock Verso Books.

\bibitem[Moynahan, 2018]{moynahan2018habermas}
Moynahan, G. (2018).
\newblock The habermas/luhmann controversy and the “cybernetics moment.
\newblock {\em Graduate Faculty Philosophy Journal}, 39(1):131--166.

\bibitem[Samuelson and Nordhaus, 2012]{samuelson2012economia}
Samuelson, P.~A. and Nordhaus, W.~D. (2012).
\newblock {\em Economics}.
\newblock AMGH Editora.

\end{thebibliography}

\end{document}